%qinvp.tex from qinv.doc T. GARAVAGLIA 
%PLAIN TEX:
\magnification 1200
\hsize = 15.truecm \vsize =22.truecm 
\hoffset= 1.0truecm \voffset  0.0truecm
\hfuzz 2.truecm
%================================TITLE PAGE================================  
{\baselineskip 12pt
\font\sstenz=cmss10 scaled \magstep0
\font\ssten=cmss10 scaled \magstep1
\def\n{\hfill\break}
%\rightline{{\sstenz hep-th/xxxxxxxxx}}
\rightline{{\sstenz qinvp.tex: dias-stp-99-11}}
\centerline{\bf INSTITI\'UID \'ARD-L\'EINN BHAILE \'ATHA CLIATH}
\centerline{(Dublin Institute for Advanced Studies, Dublin 4, Ireland)}
\vskip 24pt
%title
\centerline{\bf The quantum quasi-invariant of the time-dependent} 
\centerline{\bf nonlinear oscillator and application to betatron dynamics} 
\n\n
\centerline{T. Garavaglia$^{*\dag}$}
\n
\centerline{\it Institi\'uid \'Ard-l\'einn Bhaile \'Atha Cliath, Baile \'Atha Cliath  4, \'Eire}
\n\n
%\centerline{\bf Abstract}
\par
\vbox{\ssten 
Both the classical and quantum approximate invariants are found for  
the nonlinear time-dependent oscillator of sextupole 
transverse betatron dynamics. They are represented in terms of the elements of 
a Lie algebra associated with powers of phase space coordinates. 
The first order quantum correction to the classical quasi-invariant is found. 
} 
\noindent \n\n
%PACS: {\ssten  }\n\n
%02.20.Sv Lie Algebras
%03.65.-w Quantum mechanics
%04.25.-g Approximation methods; eqs of motion
%05.45.-a Nonlinear Dynamics
%29.27.-a Beams in accelerators
%29.27. Bd Beam dynamics
Keywords: {\sstenz nonlinear, quantum, accelerator, oscillator}
%{\sstenz Contributed paper to the XIX International Symposium on Lepton and Photon
%Interactions at High Energies, Stanford University, August 9-14, 1999.}
\vskip 120pt
\leftline{5 August 1999}
\vskip 24pt
\hrule 
\n
\item{$^*$}{\sstenz E-mail: bronco@stp.dias.ie} 
\item{$^\dagger$}{\sstenz Also Institi\'uid Teicneola\'iochta Bhaile \'Atha 
Cliath.}                                                                
\vfill
\supereject
%End of Title page.
%============================================================================

%===============================DEFINITIONS===============================
%===================Mostly not used======================================
\font\sstenz=cmss10 scaled \magstep0
\font\ssten=cmss10 scaled \magstep1

%from{qmch7.doc}
\def\a{\alpha}
\def\b{\beta}

\def\frac#1#2{{#1\over#2}}

\def\e{\epsilon}
\def\g{\gamma}

\def\lbar{{\mathchar'26\mkern-9mu \lambda}}

\def\part{\partial}

\def\too{\rightarrow}

\def\ts{\thinspace}
%======================================================================

%Using /dos/t/leabhar/qm/qmdef.doc definitions.
%Slash vectors for QED

%\def\s{\,\,\,}                                                              
%Definitions for Quantum Mechanics

\def\ah{{\hat{a}}}
\def\bh{{\hat{b}}}
\def\ad{{\hat{a}}^\dagger}
\def\bd{{\hat{b}}^\dagger}  
\def\H{{\hat H}}

\def\Ih{{\hat I}}

\def\p{{\hat p}}
\def\q{{\hat q}}

%Definitions for Operators
%Definitions for TFD.doc Thermal field dynamics.

\def\a{a} %fix this. See \ah

%Definitions for coher.doc coherent states and sqz.doc squeezed states

\def\bra#1{\langle{#1}\vert}
\def\ket#1{\vert{#1}\rangle}
\def\n{\hfill\break}

%==========================START of TEXT===============================*
%\pageno=2
%\baselineskip 20pt
%use {\rm TEXT }
{\rm
\noindent {I. INTRODUCTION}
\n
\par
Quantum effects in accelerators have been of interest for many years, 
 {[1]}, {[2]}, and {[3]}; however recently with the 
development of multi TeV colliders and interest in large linear colliders, 
they have become the subject of wide spread research {[4]}.  The 
concept of a quasi-invariant has been introduced in {[5]}, and has 
been proven useful for representing the properties of nonlinear betatron 
dynamics. The quantum version of this invariant is developed, and from it 
quantum corrections to the classical results are found. The similarity between 
the Lie algebras associated with the classical case and the quantum case are 
used to obtain the relevant results. At first the classical linear and 
classical nonlinear  cases are studied from the view point of their appropriate 
Lie algebras. These systems are quantized, and the corresponding Lie algebras 
are used to determine properties of the quantized systems. 

\par It is well known that the Courant-Snyder {[6]} invariant is 
particularly useful for determining the phase space pattern for the transverse 
dynamics of a particle in a storage ring. Using the Hamiltonian for a 
time-dependent simple harmonic oscillator, one can obtain the relevant invariant. 
However, when there are nonlinear contributions to the Hamiltonian little 
success has been achieved in finding invariants. In order to better understand 
the behavior of a particle beam, it is useful to find an approximate 
invariant, which is associated with a nonlinear time-dependent Hamiltonian. 
The method used to find the quasi-invariant for the nonlinear oscillator is  
first used in the context of classical dynamics, based on 
using the Lie algebra associated with elements obtained from powers and 
products of the position and conjugate momentum coordinates. To illustrate the 
method, an example is given for the linear system, where the invariant is 
exact, and the relevant algebra is SU(1,1) {[7]}. The method used for the linear 
system can be easily generalized to study a nonlinear one dimensional system. 
The method has the advantage that the time-dependent coefficients of the 
approximate invariant are found as the solution of a system of linear first 
order differential equations.                                                           

\par
For classical one dimensional transverse dynamics, an approximate invariant 
associated with a time-dependent Hamiltonian containing a nonlinear sextupole 
term is found. Both the Hamiltonian and the approximate invariant can be 
represented as linear sums of the elements of a Lie algebra. The invariant is 
approximate in the sense that terms of order greater than three, resulting 
from the Poisson bracket of elements of the algebra, are neglected. This 
results in a closed Lie algebra. The method is extended to quantum operators,
and a similar Lie algebra involving operator elements is found. This is used 
to obtain the quantum quasi-invariant. The relation between the 
classical result is established with the aid of coherent states associated with
the linear time-dependent oscillator.
\par
The Hamiltonian associated with nonlinear betatron dynamics studied in this 
paper is a special case of the Hamiltonian for a 
particle of mass $m$ with the one dimensional 
conventional form
$$
\hat H(t) 
=\frac{\hat p^2}{2m}
+m\omega_0^2K(t)\frac{\hat q^2}{2}+\tilde S(t)\hat q^3.      
\eqno{1.1}$$
%\eqno{1.1}$$
The method used to obtain this equation from a covariant formulation of 
storage ring dynamic is found in {[2]} and {[8]}.
This equation is put in dimensionless form
$$
H(t) =\frac{\hat p^2}{2}+K(t)\frac{\hat q^2}{2}+S(t)\hat q^3,      
\eqno{1.2}
$$
with the transformation to the dimensionless position, momentum, and energy
variables $q/q_0$, $p/p_0$,
and $H/E_0$. Here
$$
\frac{p_0^2}{mE_0}=1,\quad \frac{E_0}{m\omega_0^2}=q_0^2,\quad 
p_0q_0=\frac{E_0}{\tilde \epsilon_0}\hbar,
\eqno{1.3}
$$
with $\tilde\epsilon_0=\omega_0\hbar$.
In addition
$$
S(t)={\tilde S(t)q_0^3\over E_0}.
\eqno{1.4}$$
Introducing the dimensionless parameter
$$
\tilde\epsilon={\hbar\over p_0q_0}={\tilde\epsilon_0\over E_0}
\eqno{1.5}$$
gives for the quantum bracket of the dimensionless operators $\hat q$ and
$\hat p$
$$
[\hat q,\hat p]=i{\hbar\over p_0q_0}=i\tilde\epsilon.
\eqno{1.6}$$
This allow the results which depend upon the quantum bracket to be expressed
in terms of $\tilde\epsilon$. The quantum results, corresponding to various orders 
of $\hbar$, 
are found with $\tilde\epsilon=1$,
and associated classical results are found with $\tilde\epsilon \rightarrow 0$,
corresponding to the limit $\hbar\rightarrow 0$. For applications to 
betatron dynamics, it is conventional to use $q_0=1$ and $p_0=\vert \vec 
p\vert$, which is the magnitude of the three-momentum of a relativistic 
particle.
\n
\noindent {II. THE COURANT-SNYDER INVARIANT AND SU(1,1)}
\n
\par                                                          
The time-dependent Hamiltonian for one dimensional transverse dynamics
is written in terms of the position coordinate $q$ and the conjugate 
momentum $p$  as
$$
H(t) =\frac{p^2}{2}+K(t)\frac{q^2}{2}.
\eqno{2.1}
$$
The invariant Courant-Snyder associated with this Hamiltonian is
$$
I_0(t)=\frac{\beta(t)p^2+2\alpha(t)pq+\gamma(t)q^2}{2},
\eqno{2.2}
$$                                                                    
which satisfies the partial differential equation
$$
\frac{dI_0(t)}{dt}=\frac{\partial I_0(t)}{\partial t}+\{H(t),I_0(t)\}=0.
\eqno{2.3}
$$
The Poisson bracket of phase space functions $f(p,q)$ and $g(p,q)$ is
defined as
$$
\{f(p,q),g(p,q)\}=\frac{\partial f(p,q)}{\partial p}
\frac{\partial g(p,q)}{\partial q} -\frac{\partial f(p,q)}{\partial q}
\frac{\partial g(p,q)}{\partial p}.
\eqno{2.4}
$$
The functions $\alpha(t),\,\,\beta(t)$, and $\gamma(t)$ satisfy the
equations
$$\eqalign{
\frac{d\alpha(t)}{dt}&=K(t)\beta(t)-\gamma(t)\cr
\frac{d\beta(t)}{dt}&=-2\alpha(t)\cr
\frac{d\gamma(t)}{dt}&=2K(t)\alpha(t),\cr
}
\eqno{2.5}
$$           
where
$$
\gamma(t)=\frac{1+\alpha^2(t)}{\beta(t)}.
\eqno{2.6}
$$

\par Both the Hamiltonian Eq. (2.1) and the invariant Eq. (2.2) may be expressed
in terms of the elements of the Lie algebra SU(1,1). If one introduces the
coordinates
$$\eqalign{
a&=\frac{q+ip}{\sqrt{2}}\cr
a^*&=\frac{q-ip}{\sqrt{2}},\cr}
\eqno{2.7}
$$
with Poisson bracket
$$
\{a,a^*\}=i,
\eqno{2.8}
$$
then the functions 
$$
A_1=a^{2},\,\,\,\,A_2=a^{*2},\,\,\,\,A_3=a^*a,
\eqno{2.9}
$$
satisfy the Lie algebra of SU(1,1). Namely,
$$
\eqalign{
\{A_1,A_2\}&=4iA_3 \cr 
\{A_1,A_3\}&=2iA_1 \cr
\{A_2,A_3\}&=-2iA_2. \cr}
\eqno{2.10}
$$
In terms of the elements of the algebra Eq. (2.10), the Hamiltonian and the 
invariant become
$$
\eqalign{
H(t)&=\alpha_1(t)A_1+ \alpha_2(t)A_2+\alpha_3(t)A_3 \cr
I_0(t)&=\beta_1(t)A_1+ \beta_2(t)A_2+\beta_3(t)A_3. \cr}
\eqno{2.11}
$$
Requiring $I_0(t)$ to be real gives the relations
$$
\eqalign{
\beta_1(t)&=\beta_2^*(t)\cr
\beta_3(t)&=\beta_3^*(t).\cr}
\eqno{2.12}
$$                              
When these are substituted into Eq. (2.3), one finds, using Eq. (2.10), the
set of linear differential equations
%$$\eqalign{
%\frac{d\beta_1(t)}{dt} &+2i(\alpha_1(t)\beta_3(t)-\alpha_3(t)\beta_1(t))=0 \cr
%\frac{d\beta_2(t)}{dt} &-2i(\alpha_2(t)\beta_3(t)-\alpha_3(t)\beta_2(t))=0 \cr
%\frac{d\beta_3(t)}{dt} &+4i(\alpha_1(t)\beta_2(t)-\alpha_2(t)\beta_1(t))=0.\cr}
%\eqno{2.13}
%$$
$$ 
\left(\matrix{\frac{d \beta_1(t)}{dt}\cr 
              \frac{d \beta_2(t)}{dt}\cr
              \frac{d \beta_3(t)}{dt}\cr}\right)=  
\left(\matrix{2i\alpha_3(t)&0&-2i\alpha_1(t)\cr
              0&-2i\alpha_3(t)&2i\alpha_1(t)\cr
              4i\alpha_1(t)&-4i\alpha_1(t)&0\cr}\right)
\left(\matrix{ \beta_1(t)\cr
               \beta_2(t)\cr
               \beta_3(t)\cr}\right).                                                                       
\eqno{2.13}$$
The functions $\alpha_i(t)$ and $\beta_i(t)$ satisfy the relations
$$
\eqalign{
\alpha_1(t)=\alpha_2(t)&=\frac{K(t)-1}{4} \cr
\alpha_3(t)&=\frac{K(t)+1}{2} \cr}
\eqno{2.14}
$$
and
$$
\eqalign{
\beta_1(t)&=\frac{i\alpha(t)-\frac{\gamma(t)-\beta(t)}{2}}{2} \cr
\beta_2(t)&=\frac{i\alpha(t)+\frac{\gamma(t)-\beta(t)}{2}}{2} \cr
\beta_3(t)&=\frac{\beta(t)+\gamma(t)}{2}. \cr}
\eqno{2.15}
$$
These relations can be used to show that the system of linear differential
equations Eq. (2.13) is equivalent to the system Eq. (2.5). 
With initial values given for $\beta(t)$ and $d\beta(t)/dt$, the system
of equations Eq. (2.13) can be integrated numerically, using  Eq. (2.14) 
and Eq. (2.15).
\n
\noindent {III. THE NONLINEAR SEXTUPOLE SYSTEM}
\n
\par
Next the method described above is extended to the classical case 
when a nonlinear term
is added to the linear system Hamiltonian. However, in this case 
an approximation is made to obtain
a finite closed Lie algebra which contains seven elements.
As an example, one considers the Hamiltonian Eq. (1.2) 
where $S(t)$ is the strength of the sextupole term {[9]}.
Now defining functions of $a$ and $a^*$ as
$$\eqalign
           {A_1&=a^2,\hskip 55ptA_2=a^{*2}\cr
           A_3&=a^*a,\hskip 50pt A_4=a^3\,\,\,\,\cr
           A_5&=a^{*3},\hskip 50ptA_6=a^2a^*\cr
           A_7&=a^{*2}a,\cr}
\eqno{3.1}          
$$
one finds, keeping terms of order less than four in $a$ and $a^*$, the 
closed Lie algebra
$$\eqalign{
\{A_1,A_2\}&=4iA_3\hskip 50pt \{A_1,A_3\}=2iA_1,\cr
\{A_1,A_4\}&=0,\hskip 65pt\{A_1,A_5\}=6iA_7,\cr
\{A_1,A_6\}&=2iA_4,\hskip 47pt\{A_1,A_7\}=4iA_6,\cr
\{A_2,A_3\}&=-2iA_2,\hskip 38pt\{A_2,A_4\}=-6iA_6,\cr
\{A_2,A_5\}&=0,\hskip 65pt\{A_2,A_6\}=-4iA_7,\cr
\{A_2,A_7\}&=-2iA_5\cr
\{A_3,A_4\}&=-3iA_4,\hskip 40pt\{A_3,A_5\}= 3iA_5,\cr
\{A_3,A_6\}&=-iA_6,\hskip 45pt\{A_3,A_7\}=iA_7,\cr 
\{A_4,A_5\}&=0, \cr
\{A_4,A_6\}&=0,\hskip 65pt\{A_4,A_7\}=0,\cr
\{A_5,A_6\}&=0,\hskip 65pt\{A_5,A_7\}=0,\cr
\{A_6,A_7\}&=0.\cr}
\eqno{3.2}
$$
\par
The Hamiltonian Eq. (1.2) may be written in the form
$$
H(t)=\sum_{i=1}^7\alpha_i(t)A_i,
\eqno{3.3}
$$
where $\alpha_i(t)$, $i=1\to3$ are given by Eq. (2.14), and
$$\eqalign{
\alpha_4(t)&=\alpha_5(t)= \frac{\sqrt{2}S(t)}{4}\cr
\alpha_6(t)&=\alpha_7(t)=3\alpha_4(t). \cr}
\eqno{3.4}
$$
\par One can now find an approximate time-invariant associated with  
the Hamiltonian Eq. (3.3). This is assumed to be of the form
$$
I(t)=\sum_{j=1}^7\beta_j(t)A_j,
\eqno{3.5}
$$
which contains terms up to third order in $a$ and $a^*$.
Since $I(t)$ must be real, one finds
$$
\eqalign{
\beta_1(t)&=\beta_2^*(t),\,\,\,\,\beta_3(t)=\beta_3^*(t)\cr
\beta_4(t)&=\beta_5^*(t),\,\,\,\,\beta_6(t)=\beta_7^*(t).\cr}
\eqno{3.6}
$$
When this  along with the Hamiltonian Eq. (3.3) is substituted into
Eq. (2.3), one finds, using the algebra Eq. (3.2), the system of linear 
first order differential equations
$$
\frac{d{\vec\beta}(t)}{dt}={\bf M}(t){\vec\beta}(t), 
\eqno{3.7}
$$
where
$$
{\vec\beta}(t)= \left(\matrix{ \beta_1(t)\cr
               \beta_2(t)\cr
               \beta_3(t)\cr
               \beta_4(t)\cr
               \beta_5(t)\cr
               \beta_6(t)\cr
               \beta_7(t)\cr}\right)    
\eqno{3.8}
$$
and
{  
$$\eqalign{
&{\bf M}(t)=\cr
&\left(\matrix 
{2i\alpha_3(t)&0&-2i\alpha_1(t)&0&0&0&0\cr
 0&-2i\alpha_3(t)&2i\alpha_2(t)&0&0&0&0\cr
 -4i\alpha_2(t)&4i\alpha_1(t)&0&0&0&0&0\cr
 2i\alpha_6(t)&0&-3i\alpha_4(t)&3i\alpha_3(t)&0&-2i\alpha_1(t)&0\cr
 0&-2i\alpha_7(t)&3i\alpha_5(t)&0&-3i\alpha_3(t)&0&2i\alpha_2(t)\cr
 4i\alpha_7(t)&-6i\alpha_4(t)&-i\alpha_6(t)&6i\alpha_2(t)&0&i\alpha_3(t)&-
4i\alpha_1(t)\cr
 6i\alpha_5(t)&-4i\alpha_6(t)&i\alpha_7(t)&0&-6i\alpha_1(t)&4i\alpha_2(t)&
-i\alpha_3(t)\cr
}\right). \cr}
\eqno{3.9}
$$}                                                                            
In these expressions the $\alpha(t)^{'s}$ are given in Eq. (2.14) and
Eq. (3.4). The first three $\beta(t)^{'s}$ are given in Eq. (2.15). 
The remaining $\beta(t)^{'s}$ are found as solutions to a system of first
order differential equations Eq. (3.7). Using Eq. (3.5), the quasi-invariant may be
written in the form
$$
I=\sim I_0(t)+2\Re(\beta_4(t)A_4+\beta_6(t)A_6).
\eqno{3.10}
$$
The first term $I_0(t)$ is the function Eq. (2.2), which is an invariant for the
linear system. The remaining term may be expressed in the form
$$
c_1(t)q^3+c_2(t)q^2p+c_3(t)qp^2+c_4(t)p^3,
$$
with
$$\eqalign{
\sqrt{2}c_1(t)&=\Re\beta_4(t)+\Re\beta_6(t)\cr
\sqrt{2}c_2(t)&=-(3\Im\beta_4(t)+\Im\beta_6(t))\cr
\sqrt{2}c_3(t)&=-(3\Re\beta_4(t)-\Re\beta_6(t))\cr
\sqrt{2}c_4(t)&=\Im\beta_4(t)-\Im\beta_6(t).\cr}
\eqno{3.11}
$$
The functions $c_i(t)$, $i=1\to4$, satisfy the following system of
first order differential equations:
$$\eqalign{
{\dot c}_1(t)=K(t)c_2(t)+3S(t)\alpha(t)\cr
{\dot c}_2(t)=-3c_1(t)+2K(t)c_3(t)+3S(t)\beta(t)\cr
{\dot c}_3(t)=-2c_2(t)+3K(t)c_4(t)\cr
{\dot c}_4(t)=-c_3(t),\cr}
\eqno{3.12}
$$
where dot denotes differentiation with respect to t.
\n
\noindent {IV. THE QUANTUM LINEAR SYSTEM}
\n
%\rightline{qmch7.doc}
\par
As a first approximation for finding the quantum limits associated with the
Hamiltonian Eq. (1.2) for transverse betatron oscillations, 
one neglects the nonlinear
multipole contributions and considers for each transverse degree of 
freedom a time-dependent harmonic oscillator with Hamiltonian
$${\H}(t) = \frac{{\p}^2+K({t}){\q}^2}{2},
\eqno{4.1}$$
where ${t}$ $(c=1)$ represents arc length along an  ideal  storage ring
orbit.
The dynamical evolution of the conjugate quantum operators 
${\p}={\dot\q}$ and ${\q}$ is determined from the Heisenberg equations  
$$
\eqalign{
 \frac{d{\q}}{d{t}} &=   i[{\H},{\q}]\cr
 \frac{d{\p}}{d{t}} &=  i[{\H},{\p}].\cr}
\eqno{4.2}$$
The Courant-Snyder invariant as a function of the quantum operators $\q$
and $\p$ takes the form 
$$
2{\Ih}_0({t}) = [(w{\p}- {\dot w}{\q})^2+({\q}/w)^2].
\eqno{4.3}$$ 
The invariance follows from
$$
\frac{d{\Ih}_0(t)}{d{t}} = i[{\H}(t),{\Ih}_0(t)] + \frac{\part {\Ih}_0(t)}
{\part {t}},
\eqno{4.4}$$
along with the conditions
$$
\eqalign{
{\ddot w} + K({t})w -\frac{1}{w^3} &=0\cr
{\ddot\q} +K({t}){\q} &=0.\cr}
\eqno{4.5}$$
Expressed in the usual Courant-Synder parameters, one finds
for each transverse coordinate $\q$
$$
2{\Ih}_0= \g({t}) {\q}^2 + \a({t}) 
({\q} {\p}+{\p} {\q}) + \b({t}) {\p}^2,        
\eqno{4.6}$$
with
$$
\eqalign{
\a({t}) &= -w{\dot w}\cr
\b({t}) &= w^2\cr
\g({t}) &=\frac{1+\a^2(t)}{\b({t})}.\cr
}
\eqno{4.7}$$
\par The quantum states for this system can be constructed with the aid of
the squeezing operator {[10]} and {[11]} defined as
$${\hat S=
e^{\raise1pt\hbox{$^{\frac12(\xi^* {\ah}^2 -\xi {\ah}^{\dagger 2})}$}}},
\eqno{4.8}$$
with complex $\xi =|\xi|exp(i\phi)$ and boson operators ${\ah}$ 
and ${\ad}.$ 
The time-independent rationalized Hamiltonian is
$$
{\H}_o = \frac{{\p}^2 + {\q}^2}{2}=\ad\ah+\frac{1}{2},
\eqno{4.9}$$
where the boson operators $\ah$ and $\ad$ are found from
$$
{\q}=\frac{{\ah}+{\ad}}{\sqrt 2} \qquad {\p}
=\frac{{\ah}-{\ad}}{\sqrt 2 i},
\eqno{4.10}$$
with the commutation relations
$$
[{\q},{\p}] = i\quad\Longrightarrow\quad [{\ah},{\ad}]=1.
\eqno{4.11}$$                                               
The Courant-Synder invariant Eq. (4.3) or Eq. (4.6) is found from the time-independent
Hamiltonian using the squeezing operator Eq. (4.8) to write 
$$
{\hat I}_0({t}) = {\hat S}{\H}_o{\hat S}^\dagger 
=\left({\bd}{\bh}+\frac12\right),
\eqno{4.12}$$
where
$$
{\bh}({t})={\hat S}e^{i\theta}{\ah}{\hat S}^\dagger = 
\frac12\left(
\frac1w + w-i{\dot w}\right){\ah} +
\frac12\left(
\frac1w - w-i{\dot w}\right){\ad},
\eqno{4.13}
$$
with
$$\eqalign{
\cosh|\xi|&=\frac{1}{2}\sqrt{(1/w+w)^2+{\dot w}^2}\cr
\tan\theta&=-\frac{w{\dot w}}{1+w^2}\cr
\tan(\theta+\phi)&=-\frac{w{\dot w}}{1-w^2}.\cr}
\eqno{4.14}$$
The eigenstates of ${\hat I}_0({t})$ satisfy the eigenvalue equation
$$
\eqalign{
{\Ih}_0({t}) \ket{n,{t}} &= \left(n+\frac12\right) 
\ket{n,{t}}\cr
\ket{n,{t}} &= \frac{({\bd})^n}{\sqrt n!} \ket{0}. \cr}
\eqno{4.15}
$$
The states $\ket{n,t}$ are not Schr\"odinger states, for they 
are not solutions of the time-dependent Schr\"odinger equation
$$
i\frac{\part}{\part {t}} \ket{n,{t}}_s 
= {\H}({t}) \ket{n,{t}}_s.
\eqno{4.16}$$
However, the Schr\"{o}dinger states are of the form {[12]}
$$
\ket{n,{t}}_s = e\raise2pt\hbox{$^{i\a_n ({t})}$} \ket{n,{t}},
\eqno{4.17}
$$                                    
where the phase, as shown in Appendix A, is
$$
\a_n({t}) = -\left(n+\frac12\right) \,\int^{t} \frac{d{t'}}{w^2({t'})}.
\eqno{4.18}$$
\par
To evaluate the quantum correction to $\hat I_0(t)$ and to find the 
uncertainties associated with the operators
${\q}({t})$ and ${\p}({t})$, one must use the appropriate 
coherent state associated with ${\H}({t}).$ This state is the 
time-dependent generalization of the coherent state {[13]} obtained from the 
eigenstates of the time-independent Hamiltonian Eq. (4.9).
This is the nearest quantum 
state to the classical state of the simple harmonic oscillator. 
The coherent state for a time-dependent
simple harmonic oscillator
can be generated from the squeezed                             
ground state as
$$\ket{\beta,{t}}_s={\bf D}(\beta)\ket{0,{t}}_s, 
\eqno{4.19}$$ 
where the displacement operator ${\bf D}(\beta)$ is defined as
$$
{\bf D}(\beta) = e^{\beta {\bd}({t}) 
- \beta^* {\bh}({t})}.
\eqno{4.20}$$
Here $\beta$ is a complex parameter,
which is the eigenvalue of the operator ${\bh}({t})$. This parameter
is related to the classical value of the invariant $I_0({t})$ since
$$_s\bra{\beta,{t}}{\Ih}_0({t})\ket{\beta,{t}}_s=(|\beta|^2+1/2)
\hbar/|{\vec p}|=I_0(t)+\hbar/2|{\vec p}|. 
\eqno{4.21}$$
This includes the quantum correction $(1/2)(\hbar/|{\vec p}|)$.
The variance of an operator ${\q}$ is defined as
$$
\sigma^2(q)={_s}\bra{\beta,{t}}({\q}-\bar q)^2\ket{\beta,{t}}_s, 
\eqno{4.22}$$
where the mean value of the operator ${\q}$ is
$$
\bar q={_s}\bra{\beta,{t}} {\q} \ket{\beta,{t}}_s.
\eqno{4.23}$$
 Time-independent Hamiltonian Eq. (4.9) results are found using the 
coherent state $\ket{a}$, defined for the complex parameter $a$ as
$$
\ket{a}={\bf D}(a)\ket{a} = e^{a{\ad} 
- a^* {\ah}}\ket{0},
\eqno{4.24}$$
where the parameter $a$ is related to the classical coordinates of position 
$q$ and momentum $p$ as in Eq. (2.7).
They are
$$\eqalign{
\sigma(q)=\sigma(p)&=\sqrt{\frac{\hbar}{2}}\cr
\sigma(p)\sigma(q)&=\frac{\hbar}{2},\cr
}
\eqno{4.25}$$                                                      
which yield the minimum value for the uncertainty product. For the Hamiltonian 
Eq. (4.9), the coherent state $\ket{a}$ represents the  quantum state nearest to the
classical state, $\hbar \rightarrow 0$, for which  $\sigma(q)$, $\sigma(p)$, and
the uncertainty product are zero.
\par One can now use the states Eq. (4.19) and the definition of the variance 
Eq. (4.22) to obtain results appropriate for the a particle collider. 
For the scaling transformations frequently used in betatron dynamics
$$
\eqalign{
p &\too\frac{p}{|{\vec p}|}\cr
H &\too\frac{H}{|{\vec p}|}\cr
\hbar &\too\frac{\hbar}{|{\vec p}|},\cr
}
\eqno{4.26}$$
where the three-momentum magnitude $|{\vec p}|$ is 
$$
|{\vec p}| \approx \frac{{\cal E}}{c} 
\eqno{4.27}
$$
with relativistic particle energy ${\cal E}$, one finds that
the uncertainties and the uncertainty product, 
represented in terms of the Courant-Snyder parameters 
Eq. (4.7), are
$$
\eqalign{
\sigma(q) &=\sqrt{\frac{\hbar\b(t)}{2|{\vec p}|}} \cr
\sigma\left(\frac{p}{{\vec p}}\right) &=
\sqrt{\frac{\hbar\g({t})}{2|{\vec p}|}} \cr 
\frac{p}{|{\vec p}|}&=\frac{dq}{d{t}}\cr
\sigma(q)\sigma\left(\frac{dq}{d{t}}\right)&=
\frac{\hbar\sqrt{\b({t})\g({t})}}{2|{\vec p}|}. \cr
}\eqno{4.28}
$$
Writing the amplitude as $q_{amp}=\sqrt{(\epsilon_0/\pi)\b({t})}$ with
the emittance $\epsilon_0=2\pi I_0({t})$,
one finds the results
$$
\eqalign{
\frac{\sigma(q_{ amp})}{q_{ amp}} &=
\left(\frac{\epsilon_q}{\epsilon_0}\right)^{1/2}\cr
\e_q/\pi &={\frac{\hbar}{2|{\vec p}|}\approx
\frac{\hbar c}{2{\cal E}}\approx\frac{\lbar_{particle}}{2}},\cr}
\eqno{4.29}$$
where $\epsilon_q/\pi$, the quantum emittance, represents half the 
 resolution distance of a particle in the beam.
With the approximations
$$\eqalign{
\hbar c &\approx 2\times 10^{-19}\,{\rm TeV\ts m}\cr
{\cal E} &\approx 2{\rm TeV},\cr}
\eqno{4.30}$$
one finds
$$\e_q/\pi  \approx 5\times 10^{-20}\,{\rm m}. 
\eqno{4.31}$$
For a typical proton collider 
with $\b({t}) \approx 300\, {\rm m}$ and with $q_{ amp} \approx 3.5$ mm, 
one finds
$$\eqalign{
\epsilon_0/\pi& \approx 4\times 10^{-8}\,{\rm m}\cr
\sigma(q_{ amp})& \approx 3.9\times 10^{-6}\,{\rm mm}.\cr
}
\eqno{4.32}$$
Similarly, the angular uncertainty is
$$
\sigma\left(\frac{dq}{d{t}}\right) \approx 1.3\times 10^{-11}\, {\rm rad}.
\eqno{4.33}$$                                    
\n
\noindent {V. THE QUANTUM NONLINEAR SYSTEM}
\n
\par
The method described can be extended to the case when a nonlinear term
is added to the quantum Hamiltonian. 
As an example, one considers the quantum operator  
Hamiltonian Eq. (1.2). 
Defining operator elements  of $\ah$ and $\ad$, with $[\ah,\ad]
=\tilde\epsilon$, as
$$\eqalign
           {\hat A_1&=\ah^2\hskip 85pt\hat A_2=\ah^{\dagger 2}\cr
           \hat A_3&=(\ad \ah+\ah\ad)/2\hskip 20pt \hat A_4=\ah^3\,\,\,\,\cr
           \hat A_5&=\a^{\dagger 3}\hskip 80pt\hat 
A_6=(\ah^2\ad+\ah\ad\ah+\ad\ah^2)/3\cr
           \hat A_7&=(\ah^{\dagger 2}\ah+\ad\ah\ad+\ah\ah^{\dagger 2})/3,\cr}
\eqno{5.1}          
$$
one finds, keeping terms of order less than four in $\ah$ and $\ad$ and first 
order in the quantum parameter $\tilde\epsilon$, the 
closed approximate Lie algebra
$$\eqalign{
\lbrack\hat A_1,\hat A_2\rbrack&=4\tilde\epsilon\hat A_3\hskip 50pt \lbrack\hat A_1,\hat A_3\rbrack=2\tilde\epsilon\hat A_1,\cr
\lbrack\hat A_1,\hat A_4\rbrack&=0,\hskip 65pt\lbrack\hat A_1,\hat A_5\rbrack=6\tilde\epsilon\hat A_7,\cr
\lbrack\hat A_1,\hat A_6\rbrack&=2\tilde\epsilon\hat A_4,\hskip 45pt\lbrack\hat A_1,\hat A_7\rbrack=4\tilde\epsilon\hat A_6,\cr
\lbrack\hat A_2,\hat A_3\rbrack&=-2\tilde\epsilon\hat A_2,\hskip 35pt\lbrack\hat A_2,\hat A_4\rbrack=-6\tilde\epsilon\hat A_6,\cr
\lbrack\hat A_2,\hat A_5\rbrack&=0,\hskip 65pt\lbrack\hat A_2,\hat A_6\rbrack=-4\tilde\epsilon\hat A_7,\cr
\lbrack\hat A_2,\hat A_7\rbrack&=-2\tilde\epsilon\hat A_5\cr
\lbrack\hat A_3,\hat A_4\rbrack&=-3\tilde\epsilon\hat A_4,\hskip 40pt\lbrack\hat A_3,\hat A_5\rbrack= 3\tilde\epsilon\hat A_5,\cr
\lbrack\hat A_3,\hat A_6\rbrack&=-\tilde\epsilon\hat A_6,\hskip 45pt\lbrack\hat A_3,\hat A_7\rbrack=\tilde\epsilon\hat A_7,\cr 
\lbrack\hat A_4,\hat A_5\rbrack&=0, \cr
\lbrack\hat A_4,\hat A_6\rbrack&=0,\hskip 65pt\lbrack\hat A_4,\hat A_7\rbrack=0,\cr
\lbrack\hat A_5,\hat A_6\rbrack&=0,\hskip 65pt\lbrack\hat A_5,\hat A_7\rbrack=0,\cr
\lbrack\hat A_6,\hat A_7\rbrack&=0.\cr}
\eqno{5.2}
$$
These algebraic relations are the same as those associated with the Poisson 
bracket relations. The hermitian Hamiltonian operator and the hermitian 
quasi-invariant operator 
are
$$
\hat H(t)=\sum_{i=1}^7\alpha_i(t)\hat A_i,
\eqno{5.3}
$$
and
$$
\hat I(t)=\sum_{j=1}^7\beta_j(t)\hat A_j,
\eqno{5.4}
$$
where the $\alpha_i(t)$ and $\beta_j(t)$ are defined as before.
The equation which must be satisfied by the hermitian quasi-invariant operator
is
$$
\frac{d\hat I(t)}{dt}
=\frac{\partial \hat I(t)}{\partial t}+i\lbrack\hat H(t),\hat I(t)\rbrack=0.
\eqno{5.5}
$$
This equation along with the quantum algebra leads to the same 
set of differential equations Eq. (3.7) that appear in the 
classical case.
\n
\noindent {VI. QUANTUM CORRECTIONS} %from /dos/t/leabhar/qm/qmh.doc
%See /reduce/aoinc4.
\n
\par
The quantum corrections to the quasi-invariant are obtained using first order 
perturbation theory. 
The Boson operators which occur in the linear invariant Eq. (4.12) can be written
as
$$
\pmatrix{\bh\cr \bd\cr}=\pmatrix{f_1(w)&f_2(w)\cr 
f_2^*(w)&f_1^*(w)\cr}
\pmatrix{\ah\cr \ad\cr}.
\eqno{6.1}$$
The inverse transformation is
$$
\pmatrix{\ah\cr \ad\cr}=\pmatrix{f_1^*(w)&-f_2(w)\cr 
-f_2^*(w)&f_1(w)\cr}
\pmatrix{\bh\cr \bd\cr},
\eqno{6.2}$$
where
$$\eqalign{
f_1(w)&=\frac12\left(\frac1w + w-i{\dot w}\right)\cr
f_2(w)&=\frac12\left(\frac1w - w-i{\dot w}\right),\cr}
\eqno{6.3}$$
and
$$
\vert f_1(w)\vert^2-\vert f_2(w)\vert^2=1.
\eqno{6.4}$$
These scaled Boson operators satisfy the commutation relations
$$\eqalign{
[\ah,\ad]&=1\cr
[\bh,\bd]&=1.\cr}
\eqno{6.5}$$
\par
The operators $\q$ and $\p$ become
$$\eqalign{
\q&=(w/\sqrt{2})(\bh+\bd)\cr
\p&=(1/iw\sqrt{2})(\bh-\bd)+(\dot w/\sqrt{2})(\bh+\bd).\cr}
\eqno{6.6}$$

\par
The quantum corrections to the quasi-invariant are found from the operator
$$
\hat{I}({t}) =\hat{I}_0({t})+\hat{I}_1({t}) 
\eqno{6.7}$$
where from Eq. (4.12)
$$
\hat{I}_0({t}) = (1/2)[(w{\p}- {\dot w}{\q})^2+({\q}/w)^2],
\eqno{6.8}$$
and
$$\eqalign{
\hat{I}_1(t)=&c_1(t)\q^3+c_2(t)\frac{1}{3}(\q^2\p+\q\p\q+\p\q^2)+\cr
&c_3(t)\frac{1}{3}(\p^2\q+\p\q\p+\q\p^2)+c_4(t)\p^3.\cr}
\eqno{6.9}$$
The classical values of the operators $\q$ and $\p$ are
$$\eqalign{
\bar q&=_s\bra{\beta,{t}} {\q} \ket{\beta,{t}}_s\cr
\bar p&=_s\bra{\beta,{t}} {\p} \ket{\beta,{t}}_s.\cr}
\eqno{6.10}$$
The expectation value of the quasi-invariant operator is
$$
_s\bra{\beta,{t}}{\hat I}({t})\ket{\beta,{t}}_s=
_s\bra{\beta,{t}}{\hat I}_0({t})\ket{\beta,{t}}_s+
_s\bra{\beta,{t}}{\hat I}_1({t})\ket{\beta,{t}}_s,
\eqno{6.11}$$
where $_s\bra{\beta,{t}}{\hat I}_0({t})\ket{\beta,{t}}_s$ is given in 
Eq. (4.21).
The correction to the linear invariant is
$$
_s\bra{\beta,{t}}{\hat I}_1({t})\ket{\beta,{t}}_s=I_1+I_{1qc}.
\eqno{6.12}$$
The classical correction to the linear invariant is
$$
I_1=c_1(t){\bar q}^3+ c_2(t){\bar q}^2{\bar p}
+c_3(t){\bar p}^2{\bar q}+c_4(t){\bar p}^3,
\eqno{6.13}$$
and the quantum correction is
$$\eqalign{
I_{1qc}&=((\bar{q}/2)[\beta(t)c_1(t) 
+ \gamma(t)c_3(t) - \alpha(t)c_2(t)]\cr
&+ (\bar{p}/2)[\beta(t)c_2(t) 
+  3\gamma(t)c_4(t) - \alpha(t)c_3(t)])(\hbar/\vert\vec p\vert).\cr}
\eqno{6.14}$$
\n
\noindent {VII. RESULTS AND CONCLUSIONS}
\n
\par                                                          
The nonlinear time-dependent Hamiltonian for one dimensional 
transverse classical dynamics
is written in terms of the position coordinate $q$ and the conjugate 
momentum $p$  in Eq. (1.2).
For this Hamiltonian, the equation of motion is found from Hamilton's equations
$$\eqalign{
{\dot q}&=\frac{\partial H(t)}{\partial p}\cr
{\dot p}&=-\frac{\partial H(t)}{\partial q}\cr}
\eqno{7.1}
$$
to be
$$
{\ddot q}+K(t)q+3S(t)q^2=0.
\eqno{7.2}
$$
The classical approximate invariant associated with this Hamiltonian is
$$\eqalign{
I(t)&=\frac{\beta(t)p^2+2\alpha(t)pq+\gamma(t)q^2}{2}\cr
    &+c_1(t)q^3+c_2(t)q^2p+c_3(t)qp^2+c_4(t)p^3.\cr} 
\eqno{7.3} $$                                                                    
The time-dependent functions $\alpha(t)$, $\beta(t)$,  and $\gamma(t)$
are found from the Eq. (2.5) or Eq. (2.13), and the functions $c_i(t)$ can be 
found from the differential equations Eq. (3.12). 
These system of equations are  
equivalent to the system of linear equations Eq. (3.7). 
\par  
Numerical results are given which confirm the analytical
development in the previous sections. Periodic solutions for the functions
$c_i(t)$ allow the determination of these functions at a fixed point 
in a lattice with a sextupole nonlinearity. The values of the functions
$q$ and $p$ are determined from nonlinear tracking for the first five
circuits of the lattice. After the $j^{th}$ turn, the quasi-invariant
becomes
$$
I(j)=I_0(j)+c_1(j)g(1,j)+c_2(j)g(2,j)+c_3(j)g(3,j)+c_4(4,j),
\eqno{7.4}
$$
with
$$
\eqalign{
g(1,j)&=q(j)^3\cr
g(2,j)&=q(j)^2p(j)\cr
g(3,j)&=q(j)p(j)^2\cr
g(4,j)&=p(j)^3.\cr}
\eqno{7.5}
$$
From the requirement that
$$
I(k)-I(1)=0,
\eqno{7.6}$$
for $k=2\to5$, one finds the system of linear equations
$$
\Delta(k)=\sum_{i=1}^4c_i(j)\Delta g(i,k),
\eqno{7.7}
$$
with
$$
\eqalign{
\Delta(k)&=-(I_0(k)-I_0(1))\cr
\Delta g(i,k)&=g(i,k)-g(i,1)\cr}
\eqno{7.8}
$$
For the numerical results, the FODO approximation is used to find the lattice
function $\beta(t)$. This function is derived in Appendix B.
The system Eq. (7.7) and Eq. (7.8) is solved numerically to find the coefficients
$c_i(t)$. The thin lens approximation is used where the lattice is made
up of a single thin sextupole element and identical cells of length $L$. 
Each cell consists of a 
focusing and defocusing
magnet separated by a bending drift magnet. The focal
length of the focusing and defocusing magnets is $f$, and the phase
advance per cell $\mu$ is found from
$$
\sin(\mu/2)=\frac{L}{2f}.
\eqno{7.9}
$$
The tune $\nu$ is obtained from
$$
\nu=\frac{\mu N_c}{2\pi},
\eqno{7.10}
$$
where $N_c$ is the number of cells. The maximum value of $\beta(t)$
occurs when $\alpha(t)=0$, and $\beta(t)=1/\gamma(t)$, and it is found
from 
$$
\beta(t)_{max}=2f\left(\frac{1+\sin(\mu/2)}{1-\sin(\mu/2)}\right ).
\eqno{7.11} $$ 
\par 
The phase space plot of $\beta_{max}p$ cm and $q$ cm for the 
classical quasi-invariant $I$ is shown in Figure. 1. The classical 
results for both the invariant for the linear system $I_0$ cm and the 
quasi-invariant $I$ cm for the nonlinear system are plotted in Figure 2. as a 
function of turn number. It is clearly seen here that the methods leading to 
the quasi-invariant produce a more stable quantity than $I_0$.
For the 
example considered, the values $N_c=4$, $\mu=\pi/2$, $L/2=8875\,\,$mm, and 
$\nu= 0.33666667+N_c\mu/(2\pi)$, with near resonance fractional tune contribution, 
have been used. The initial values $q=0.3$ cm, 
$\beta_{max} p=0$ along with the sextupole strength 
$3s_e=0.1\times 10^{-5}\,{\rm cm}^{-2}$ have 
been used. For integer $j$, the sextupole function is approximated 
by $S(t)=(s_e/3)\delta(t-jT_0)$, where 
$T_0$ is the orbital period.
For the present case, the
values of the periodic functions $c_i(0)$ are
\vbox{ 
$$
c_1= -3.41219\times10^{-6}\,\,\,\hbox{\rm cm}^{-2}
$$
$$
c_2/\beta_{max}=-0.91910\times10^{-7}\,\,\,\hbox{\rm cm}^{-2}
$$
$$
c_3/\beta_{max}^2=+0.99563\times10^{-5}\,\,\,\hbox{\rm cm}^{-2}
$$
$$
c_4/\beta_{max}^3=-1.11468\times10^{-7}\,\,\,\hbox{\rm cm}^{-2},
$$
$$
\eqno(7.12)$$
} 
with $\beta_{max}=38389.279$ cm.
\par 
It is clear from the Figure 2. that the quasi-invariant is nearly 
stable. It remains this way for increasingly larger number of turns. It 
oscillates with small amplitude and with period of $100$ turns. The amplitude 
of the oscillation depends upon the strength of the sextupole nonlinearity, 
and the period results from the nearness of the fractional tune to the third 
integer resonance. Although, the present quasi-invariant, which includes terms 
in $q$ and $p$ through third order, becomes increasingly unstable for large 
values of the sextupole strength or large initial values of the 
amplitude $q$, 
it is clear that the method can be extended to include arbitrarily higher 
order corrections which will improve the stability of the quasi-invariant.  
The quantum correction associated with the quasi-invariant can be found 
from Eq. (4.21) and Eq. (6.14), and for the numerical example being considered
it takes the value
$$
I_{qc}\approx (1/2+{\bar q}/2(\beta_{max}c_1+c_3/\beta_{max}))
\hbar/\vert \vec p\vert \approx 0.538\,\hbar/\vert \vec p\vert.
\eqno{7.13}$$
Although very small for a hadron collider, it would be more significant 
for a low energy nonlinear time-dependent oscillator of the type described by
the Hamiltonian Eq. (1.2).
\par
In conclusion, it is seen that the Lie algebra methods used for both the classical 
and quantum quasi-invariants provide a useful approximation for the invariant
associated with the time-dependent nonlinear oscillator. For applications to 
betatron dynamics, this method provides a complimentary method to the usual 
nonlinear-map tracking methods. In addition, the quantum states $\ket{\beta,t}_s$
of Eq. (4.19) provides the connection between the quantum operator for 
the quasi-invariant and the classical result when these states are used to 
form matrix elements of the type used to obtain the quantum uncertainties 
Eq. (4.28) and the quantum correction Eq. (6.14).
\par This work was performed in part while the author was University Scholar 
in Theoretical Physics at UCLA and partially supported by 
U.~S.~Department of Energy Contract No.~DE-AC35-89ER40486. Additional  
support came from Institi\'uid Teicneola\'iochta Bhaile \'Atha Cliath 
grant 9571. Computations have been done with the aid of REDUCE and the 
CERN Computer Library.
\vfil \eject
\noindent APPENDIX {A}: {THE SCHR{\" O}DINGER STATE PHASE}
\n
\par                                                          
The phase Eq. (4.18) can be found by first differentiating Eq. (4.17) 
with respect to ${t}$, and then using Eq. (4.16) 
to write the matrix element
$$
i\frac{\partial \a_n({t})}{\partial {t}} + \bra{n,{t}}
\frac{\partial}{\partial {t}}\ket{n,{t}}
= -i\bra{n,{t}}{\bf H}({t})\ket{n,{t}}
$$
$$=\frac{1}{2}({{\dot w}}^2+K({t})w^2+1/w^2)(n+1/2),
\eqno{A.1}$$
where Eq. (4.13) is used to express the 
Hamiltonian Eq. (4.1) as a function of
${\bf b}({t})$ and its adjoint.
The matrix elements of this operator can be found from
$$
{\bf b}^{\dagger}\ket{n,{t}} = \sqrt{n+1}\, \ket{ n+1,{t}}
\eqno{A.2}$$
and
$${\bf b}\,\ket{n,{t}} = \sqrt n \,\ket{n-1,{t}} \quad\Longrightarrow
\quad
\bra{n,{t}}{\bf b}^{\dagger} = \sqrt n \bra{n-1,{t}}.
\eqno{A.3}$$
Making the replacement $n \rightarrow n-1$ in Eq. (A.2), 
one can derive the identity
$$
\bra{n,{t}}\frac{\partial {\bf b}^{\dagger}}{\partial {t}} 
\ket{n-1,{t}} +
\bra{n,{t}}{\bf b}^{\dagger} \frac{\partial}{\partial 
{t}}\ket{n-1,{t}}
$$ 
$$=
\sqrt n \bra{n,{t}}\frac{\partial}{\partial {t}}\ket{n,{t}},
\eqno{A.4}$$
where
$$
\frac{\partial {\bf b}^{\dagger}}{\partial {t}} =
\frac{1}{2}(i(w{\ddot w}-{{\dot w}}^2)-2{\dot w}/w){\bf b} 
+ i(w{\ddot w}-{\dot w}^{2}){\bf b}^{\dagger}.
\eqno{A.5}$$
Therefore
$$ 
\bra{n,{t}}\frac{\partial}{\partial {t}}\ket{n,{t}}  
= \bra{0,{t}}\frac{\partial}{\partial {t}}
\ket{0,{t}}
+
i\,\frac{t}{2} (w {\ddot w} - {\dot w}^{2}).
\eqno{A.6}$$
Choosing
$$
\eqalign{
\bra{0,{t}}\frac{\partial}{\partial {t}}\ket{0,{t}} 
&=\frac{i(w{\ddot w}-{\dot w}^{2})}{4}\cr},
\eqno{A.7}$$
one finds from Eq. (4.5), Eq. (A.1), Eq. (A.6), and Eq. (A.7) the differential equation 
$$
\frac{d\a_n({t})}{d{t}} =
-(n+\frac12)\,\frac{1}{w^2({t})},
\eqno{A.8}$$
which has the solution Eq. (4.18).
\vfill\eject
\noindent APPENDIX {B}: {THE BETA LATTICE FUNCTION}
\n
\par In this appendix, following the methods of {[6]}, the 
lattice function $\beta(t)$ used in the numerical calculations 
is derived. It is found for a lattice made up of similar cells of the 
FODO (focusing, drift, defocusing, drift) form. Focusing and defocusing
are achieved with thin lens quadrupole magnets, and the drifts occur through
bending dipole magnets of length $L$ and strength $B_0$. The function
$\beta(t)$  
has period $2L$, and the function on the interval $L<t<2L$ is 
found from that on
the interval $0<t<L$ using
$$
f(t)_{L<t<2L}=f(2L-t)_{0<t<L}.
\eqno{B.1}
$$
\par The beta functions for a lattice with phase advance $\mu$ 
per cell are found from the $(1,2)$ component 
of the transfer matrix. The function $\beta(t)$ 
is found from 
$$\beta(t)\sin\mu=({\bf O}(t){\bf F}{\bf O}(L)
{\bf D}{\bf O}(L-t))_{12},
\eqno{B.2}
$$
where the focusing and defocusing matrices for lenses of 
focal length $f$ are, respectively,
$$
{\bf F}=\left( \matrix{1&0\cr
                       -1/f&1\cr}\right)\,\,\,\,{\rm and} 
\,\,\,\,
{\bf D}=\left( \matrix{1&0\cr
                      1/f&1\cr} \right).
\eqno{B.3}
$$
The matrix for a drift of distance $t$ is
$${\bf O(t)}= \left( \matrix{1&t\cr
                      0&1\cr} \right).
\eqno{B.4}$$
The resulting beta function for $0<t<L$ is
$$
\beta(t)=\frac{2L}{\sin\mu}\left(1+\frac{L}{2f}-(\frac{1}{f}
+\frac{L}{2f^2})t
+\frac{t^2}{2f^2}\right), 
\eqno{B.5}$$
where $L=2f\sin(\mu/2)$.
\vfill\eject
%======================TEX REFERENCES=======================================
\noindent{REFERENCES}
\vskip 12pt
\hrule
\vskip 24pt
\item{$^*$}E-mail: bronco@stp.dias.ie
\item{$^\dagger$}Also Institi\'uid Teicneola\'iochta Bhaile \'Atha Cliath. 

\item{[1.]}
A.~A.~Sokolov, and I.~M.~Ternov, 
{\it Radiation from Relativistic Electrons} 
(American Institute of Physics, New York, 1986).
 
\item{[2.]}
T.~Garavaglia, in
 {\it Conference Record of the 1991 IEEE Particle Accelerator Conference, 
San Francisco}, edited by L.~Lizama and J.~Chew (IEEE, 1991) 
Vol. I, p. 231.
 
\item{[3.]}
T.~Garavaglia, in
 {\it Proceedings of the 1993 Particle Accelerator Conference, Washington 
D.~C.} (IEEE, 1993) Vol. V, p. 3591.
 
\item{[4.]} 
 {\it Quantum Aspects of Beam Dynamics}, edited by  Pisin Chen 
(World Scientific, 1999).  
 
\item{[5.]}
T.~Garavaglia, in
 {\it Proceedings of International Conference on High Energy Physics, 
Dallas 1992, Conference Proceedings No. 272 }, edited by J.~R.~Sanford 
(American Institute of Physics, New York,  1993) Vol. II, p. 2026.
 
\item{[6.]}
 E.~D.~Courant and H.~S.~Snyder,    
 Annals of Phys. (N.~Y.) {\bf 3}, 1 (1958).
 
\item{[7.]}
A.~Perelomov, 
 {\it Generalized Coherent States and Their Applications} 
(Springer Verlag, Berlin, 1989) p. 67.     
%SU(1,1) algebra.  

\item{[8.]}
T.~Garavaglia, in
 {\it Proceedings of the 1991 Symposium on the 
Superconducting Super Collider, Corpus Christi, SSCL-SR-1213}, 
edited by V.~Kelly and G.~P.~Yost
(Superconducting Super Collider Laboratory, Dallas Texas, 1991), p. 669.
 
\item{[9.]}
E.~J.~N.~Wilson, in  {\it CERN, Proc. No. 87-03}, 
edited by S. Turner 
(CERN, Geneva, 1987) p. 57.
%Sextupole Dynamics 

\item{[10.]}
D.~F.~Walls, Phys. Rev. {\bf 306}, 141 (1983).
 
\item{[11.]}
I.~A.~Pedrosa, Phys. Rev. {\bf D36}, 1279 (1987).

\item{[12.]}
H.~R.~Lewis, Jr. and W.~B.~Riesenfeld, J.~Math.~Phys. {\bf 10}, 1458 (1969).
 
\item{[13.]}
R.~J.~Glauber, Phys. Rev. {\bf 131}, 2766 (1963).
\vfill\eject
%===========================FIGURE CAPTIONS===================================
%\vskip28pt
\noindent{FIGURE CAPTIONS}
\vskip28pt
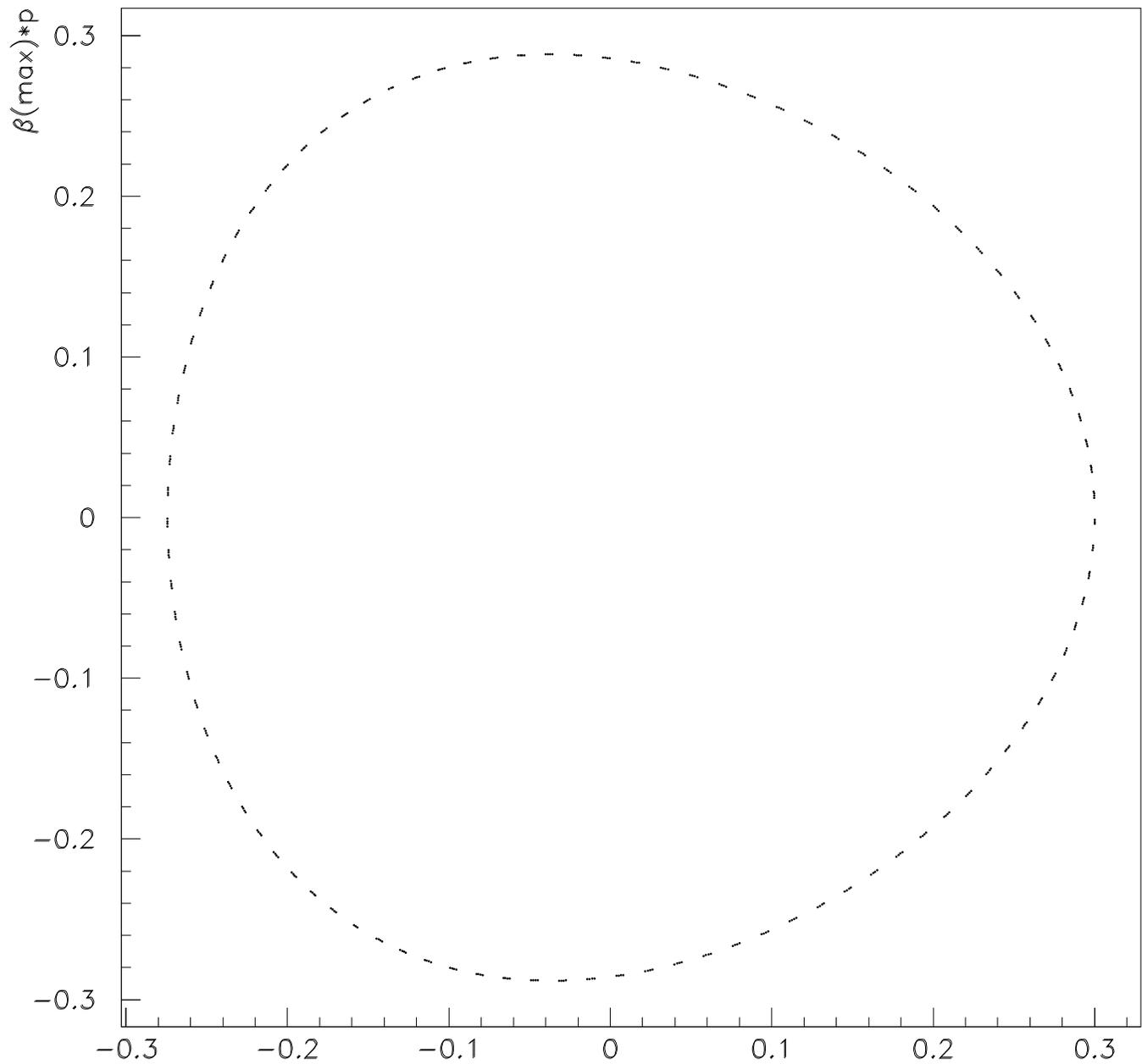
\item{FIG. 1.} The quasi-invariant in phase space using $\beta_{max}\,p$ cm and $q$ cm. 
                                                   
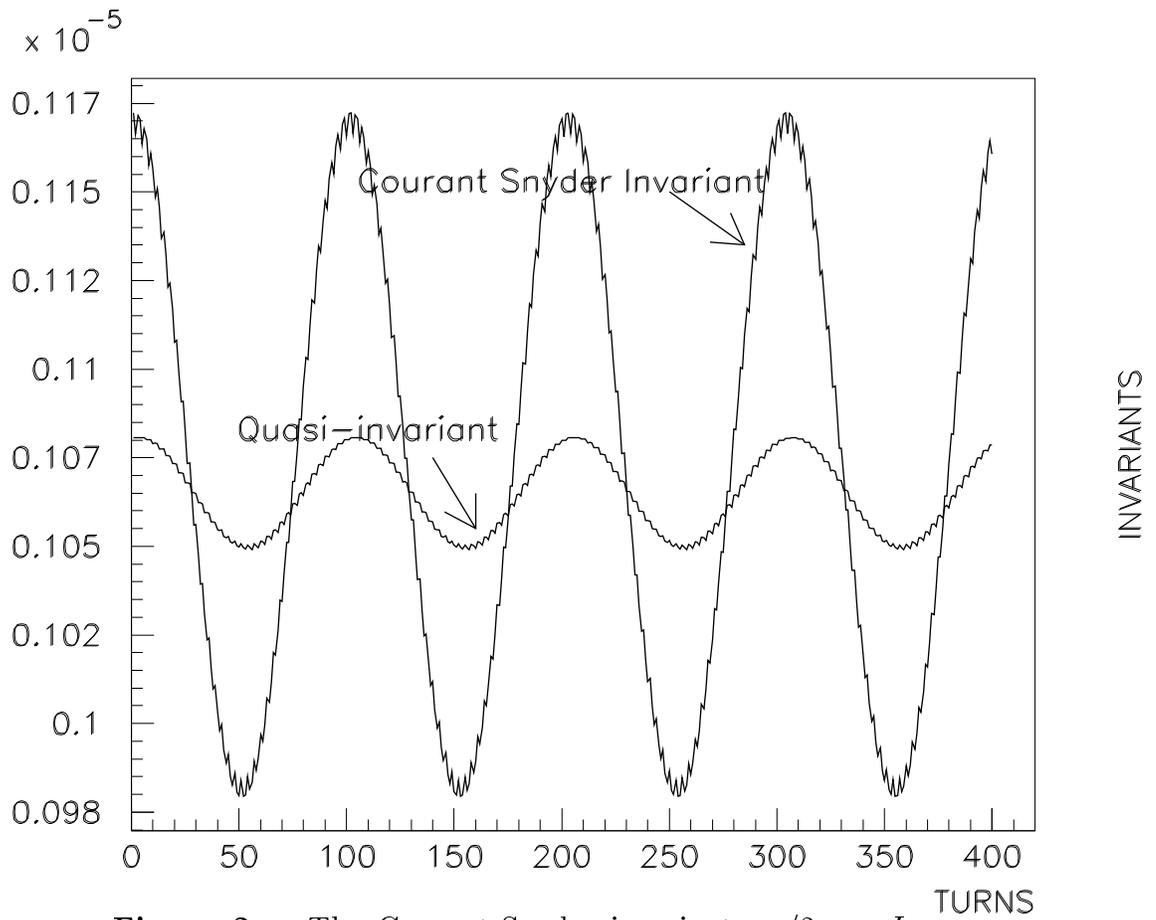
\item{FIG. 2.} The Courant-Snyder invariant, $\epsilon_0/2\pi=I_0$ cm, and the 
               quasi-invariant $\epsilon/2\pi=I$ cm, as a function of turn-number.

\vfill\supereject
%===========================END OF QINVP.TEX===================================
}
\bye